# How do Software Startups Pivot? Empirical Results from a Multiple Case Study

Sohaib Shahid Bajwa[1,3], Xiaofeng Wang[1, 3], Anh Nguven Duc [2,3], Pekka Abrahamsson[2,3]

[1] Free University of Bozen-Bolzano, Piazza Domenicani 3, 39100, Bolzano, Italy,
http://www.unibz.it
[2] Norwegian University of Science and Technology, NO-7491 Trondheim, Norway
[3] Software Startups Research Network http://softwarestartups.org

**Abstract.** In order to handle intense time pressure and survive in dynamic market, software startups have to make crucial decisions constantly on whether to change directions or stay on chosen courses, or in the terms of Lean Startup, to pivot or to persevere. The existing research and knowledge on software startup pivots are very limited. In this study, we focused on understanding the pivoting processes of software startups, and identified the triggering factors and pivot types. To achieve this, we employed a multiple case study approach, and analyzed the data obtained from four software startups. The initial findings show that different software startups make different types of pivots related to business and technology during their product development life cycle. The pivots are triggered by various factors including negative customer feedback.

**Keywords:** Software Startup, Lean Startup, Pivot, Validated Learning

## 1   Introduction

Many people know Twitter as arguably the most famous microblogging platform. Much less are aware that it was a podcast service provider back in its startup phase in 2005 [8]. Similarly, Instagram back in its early days was a social check-in application called Burbn, combining features of a photo share app (Foursquare) and a game (Mafiawars) [7]. As the examples show, very few software startups get their products or business right immediately, and most do not end up with what they had initially started.

   This is because software startups intend to produce cutting edge products and grow fast under the condition of extreme technology and business uncertainty. In order to obtain a sustainable business model, software startups change their direction relentlessly, or make a *pivot* in Lean Startup approach [1]. Ries [1] defines *pivot* as a strategic change, designed to test a fundamental hypothesis about a product, business model or engine of growth. Pivot is often considered the outcome of validated learning, another key concept of the Lean Startup to test a business hypothesis and measure the result. Software startups often neglect the validated learning process and avoid pivot when needed, which is one of the reasons behind many startup failures

[2]. Pivot is considered vital for software startups to survive, grow, and eventually obtain a suitable business model.

Due to the nascent nature of software startup research, previous empirical studies specially focusing on pivot are scarce. To the best of our knowledge, no study has been conducted exploring different types of pivots and identifying different triggering factors. This study attempts to fill this knowledge gap, examining pivots in software startups during different product development stages, from concept to mature product. The main objective of our study is to provide a better understanding of pivots happening in software startups. To this end, the main research question asked in the study is: *How do software startups pivot during different product development stages*?

The rest of this paper is organized as follows. In Section 2, background and related work is presented. Section 3 describes the empirical research approach. The findings are presented in detail in Section 4 and further discussed in Section 5. The paper is summarized in Section 6 outlining the future research.

## 2   Background and Related Work

Pivot is a core concept of Lean Startup [1], a startup methodology that focuses on the Build-Measure-Learn (BML) loop with three steps: turn idea into product, measure its effect, and learn from the result. This learning is referred to as validated learning [1]. Each hypothesis regarding the business model is tested, and a decision is made accordingly on whether to pivot or persevere. Pivot is not about introducing just any change, even though the two terms are often used as synonyms. Pivot is a special kind of change designed to test and validate the assumptions a startup has about its product, business model, and the engine of growth [1]. Ries presents ten different types of pivots that can happen in a startup [1], listed in Table 1.

**Table 1.** Pivot types [1]

| | |
|---|---|
| Zoom-in | A single feature of a product becomes the whole product itself. |
| Zoom-out | The whole product becomes a single feature of a much larger product, mainly because the original product is insufficient to address customer needs. |
| Customer segment | While trying to solve the right problem, a startup discovers a different segment of customers than originally anticipated. |
| Customer need | A startup realizes the problem they try to solve is not very important for the customers, and discovers other related problems that are more important. |
| Platform pivot | An application is turned into its supporting platform or vice versa. |
| Business Architecture | A startup switches its business architecture e.g. aiming for low volume, high margin, instead of focusing on mass market. |
| Value Capture | Changing the way/method to capture value (monetize) for a startup. |
| Engine of Growth | A startup makes significant changes in its growth strategy to seek rapid and more profitable growth. |
| Channel Pivot | A startup has identified a more effective way to reach its customers |
| | than its previous one. |
| Technology Pivot | A startup delivers the same solution by using completely different technology. |

Only few studies touch upon the topic of startup pivot [2,3,4,9,10]. By providing evidence of two real world software startup failures, Giardino et al. [2] concluded that neglecting the learning process and avoiding pivot can become the reasons of software startup failure. Bosch et al. [9] offer an alternative to pivot or persevere. They present a software development model for early stage software startups. But the study is not primarily focused on investigating pivot in software startups. Another study related to pivot was conducted by Van der Van and Bosch [4], which gives a broader overview on pivots in software startups. It compares the similarities and differences between pivot decisions and software architecture decisions. The study considers a pivot as a business decision only, and is not primarily focused on how software startups pivot during their life cycles.

The work closely related to this study was from Terho et al. [3], in which the authors explain how pivots can change business hypotheses in a lean canvas model. They have identified some pivot types, e.g. zoom-out, customer segment, and platform pivots. However, there is a lack of evidence of how they were identified and how the link between pivot types and lean canvas was built.

Based on the observed knowledge gap, we focused our study on understanding pivots in software startups by identifying their types and examining the factors triggering them. As Nguyen-Duc et al. [10] argue, different types of pivots might happen in different phases of a startup's lifecycle. Therefore we adopted a product development perspective on the phases of a startup's lifecycle. A startup goes through different product development stages during their life cycles, which are: concept, in development, working prototype, functional product with limited users, functional product with high growth, and mature product [6]. The product development stages allow us to obtain a contextualized understanding of pivots in software startups.

## 3  Research Approach

Given the exploratory nature of our study and the "how" research question, we employed a multiple qualitative case study approach [11]. The selected four cases were software-based startups. Each was covering a different product development stage at the time of our study (Table 2). All pivoted during their product development.

The main data collection method was interviews. We conducted semi-structured interviews with open-ended questions. Each interview lasted from 30 minutes to one hour, and was transcribed for further analysis. All of the interviewees were the founders, were involved in the decision making process and knew the journey of their startups from the inception till today.

The data analysis followed the multiple-case analysis suggested by Yin [11]. Within-case analysis was conducted firstly. Then in the cross-case comparison, the identified pivots and different triggering factors causing pivots across cases were compared and contrasted.

**Table 2.** Profiles of the Software Startups

| Software Startup | Business Domain | Founded | # of Founders | Current Product Dev. Stage | Country |
|---|---|---|---|---|---|
| Dicy | Video service | 2014 | 2 | Working prototype | Italy |
| DocMine | Software as a service | 2015 | 3 | Functional product with limited users | Austria |
| Hooka | Event Ticketing system | 2011 | 2 | Functional product with high growth | Norway |
| Easy Learning | Game based Learning | 2006 | 2 | Mature Product | Norway |

## 4 Results

**Case 1: Dicy**
Dicy is a software-based startup that provides video service specially designed for other startups to create their promotion videos. It started as an online community platform in 2014, where entrepreneurs could meet, share their ideas and also ask for different resources according to their needs. In July 2014, when their online platform already had limited users, they identified a different need of customers, and pivoted from online community platform towards providing video service facilities for startups.

The main factor causing this customer need pivot was feedback obtained from the customers, according to the co-founder of Dicy we interviewed: *"We realized that most of the startups, especially software startups, don't really want to talk about their ideas because of people stealing their ideas."* The co-founder explained the rationale behind conducting this pivot: *"We decided if we want to help startups in this communication, we need to find a different solution, in order to approach to investors in an easy and comfortable way."* The outcome of this pivot was positive, as the co-founder stated: *"During the trial stage, we could see that the concept was kind of approved. We see that this demand exists."*

When asked about the realization of being flexible and allowing pivot, the co-founder suggested: *"You would realize something is not coming up. And you need change. That's very important. You should allow this kind of change. You need to be flexible. You need to try out what are you capable of doing and what it takes to."*

**Case 2: DocMine**
The original idea of DocMine when it started was to develop better encryption software. They made their first significant change when they pivoted towards providing a unified API to access different social media sources, such as Facebook and Twitter. The founder commented on the reason behind this pivoting: *"In Sweden, another company is also working on this and developing better than us. So we shifted and stopped working on this idea."* When describing the reason of not competing with their competitors, the founder described: *"You have to react very fast in this IT*

*world. If you have idea, you have to react fast and take your product into market quickly especially if there is another one."* Consequently DocMine conducted different brainstorming and mind mapping sessions within the team to discover the new direction of their startup.

Meanwhile, there has been significant change in the DocMine founding team. Before the pivot described above, one of the co-founders left the team and worked for Audi. While working on the new social media API idea and the working prototype of the new product was developed, the left co-founder requested to rejoin the team. At that time DocMine already hired one person due to this co-founder's leaving. However DocMine decided to take back their co-founder. The founder explained the rationale behind this decision: *"We decided to take him [back], not only because of performance, but because we knew him. So when you are working together, you know who is he, how he thinks, I think it's very important for startup to have perfect group dynamic."*

In August 2015, DocMine discovered from the feedback of their customers that their solution seemed to be more interesting for developers rather than the private markets and different companies they initially conceived. Even though at that time the new product was already functional with limited users, they shifted their focus to the developers, therefore pivoted in terms of customer segment.

**Case 3: Hooka**
Hooka provides a ticketing system for different events focusing on small companies in a user friendly way. The main focus is on small companies who cannot afford expensive solutions to organize events. Their initial service, however, was completely different: selling magnetic cubes. The business did not get much traction and they made a complete pivot in 2011.

It is followed by another pivot related to customer segment in the concept stage of the ticketing system idea. The initial focus of the idea was to develop a bidding system for bar and nightclub seats. The founder described the situation before pivoting: *"We started to do some research and found out that discos and clubs would not want a product. They manage things fine as it was."* Due to this negative feedback from the potential customers, they pivoted towards providing a ticket validation system for small companies who organize events.

At the same time, they also made significant change in their product from providing a simple SMS-based application towards developing a complete ticket validation system. This is an example of zoom-out (product) pivot.

A team pivot was made when the prototype did not work. The founder described: *"They (developers) use Google to search for the code. They are 'copy-paste' programmers. They are not skilled enough basically. They made a prototype that barely held together... I need to hire the professionals. We did not find any previous work useful and we scrapped everything."* As a result Hooka changed the whole development team, and hired new professionals who could develop.

**Case 4: EasyLearning**
EasyLearning is a game-based learning platform to be used in the classrooms (or in any other learning environment) in which a teacher asks a quiz and students answer using their mobile devices. Initially it started with developing quiz for Sony Phones or

PC's, but later pivoted to developing quiz for iPhone and Android. The main factor causing this pivot was the emergence of the smartphone, as the co-founder explained: *"At that time, Android and iPhone was not available. It is a major change. So after maybe in 2008 or 2009 we got smartphones and tablets with proper web browsers then we could make a web-based client to make it a lot easier for development and we make the whole platform from scratch."*

Although this technology pivot solved their problem of involving a maximum number of students simultaneously, it had consequences. The co-founder recalled: *"The major issue is, due to the early web technology, due to slow, large latency, we did not implement web socket or something like that. The client was just pulling the server, the performance is horrible because client keep pulling the server all the time."*

In order to solve these issues, in the version 3.0, they threw away everything and started from scratch. The CEO explained: *"The main different is nice user interface, java based server with graphic engine. Web based client as before, but we have editor to create quiz which was not possible before."*

Table 3 provides a summary of the pivot types and triggering factors found in the four studied software startup companies.

**Table 3.** Summary of Pivots in Software Startups

| Case Name | Pivot Type | Triggering Factor | At Which Product Dev. Stage Pivot Happened |
|---|---|---|---|
| Dicy | Customer need | Negative customer feedback | Functional product with limited users |
| DocMine | Complete | Failing to compete with competitors | Functional product with limited users |
| | Team | Founder's decision | Working prototype |
| | Customer segment | Negative customer feedback | Functional product with limited users |
| Hooka | Complete | Negative customer feedback | Functional product with limited users |
| | Customer segment | Negative customer feedback | Concept |
| | Product Zoom-out | Negative customer feedback | Concept |
| | Team | Missing team competence | Working prototype |
| Easy Learning | Technology | Emergence of smartphone | Working prototype |
| | Technology | Technology limitation | Functional product with high growth |

## 5   Discussion

Software startups often lie in the soil of extreme uncertainty, and do not know their customers in advance. Although they are solving a problem, their initially perceived customers may not be interested in that problem. Startups try their ideas and learn from their failure, and pivot towards the real needs and right segments of the customers. As shown in the cases of Dicy, DocMine and Hooka, they all experienced customer need or segment pivots. Learning from their initial failures, they identified the right problems or customer segments and pivoted towards the new directions. In the cases of DocMine and Hooka, some pivots were so profound that almost all aspects of the startups were changed, product, targeted market and business model, only the original entrepreneurial team remained the same. We termed this type of pivot *complete pivot*. This is an addition to the pivot types listed in Table 1 [1].

Building an entrepreneurial team is one of the key challenges faced by many software startups [5]. As a response to this challenge, the entrepreneurial teams go through significant changes in team composition. This kind of pivot is termed as team pivot. It is another addition to Table 1. The change can be related to key members (e.g. co-founder) or having a new development team completely. Both Hooka and DocMine experienced team pivots. A lack of competency needed can be one factor causing software startup teams to pivot, as exhibited in the Hooka case. In contrast, DocMine evidences the team pivots as consequences of their founders' decision.

Our study also indicates the potential links between pivots. It happened in all but Dicy case that one pivot caused another pivot, therefore an evidence of what Terho et al. [3] call the "domino" effect. For example, after Hooka decided to make customer segment pivot, they soon realized that their original solution was only a feature of a much larger solution and therefore performed a product zoom-out pivot. Future studies need to be conducted in order to investigate the effect of different types of pivots, and the relationship among them.

The validity threats of our study are hereby discussed. One validity threat is related to the generalizability of the results. As our study is exploratory in nature, qualitative case study is a suitable approach to understand how software startups pivot in real world. More case studies and further quantitative studies need to be conducted e.g. survey, in order to make the result more generalizable. Another validity threat is related to the interviewees and their knowledge about their startups' history. It is mitigated to a large extent by interviewing the founders who generally have the best knowledge of their startup processes.

## 6   Conclusions

Software startups are developing cutting-edge software products significantly contributing to the world economy. However, in order to achieve success, most of the software startups need to learn and pivot continuously. This paper provides a deeper contextual understanding of how software startups pivot, employing a multiple qualitative case study approach.

The findings of the study show that software startups make different pivots in early product development stages. Customer segment and technology pivots are common. The pivots can be triggered by different factors. Negative feedback from customers is the major factor causing pivots.

We call for further investigation on the consequences and relationship among different pivots. Further quantitative studies (e.g. survey) need to be conducted to obtain quantitative validation and to generalize the results.

## Acknowledgement

We are thankful to Pertti Seppänen from University of Oulu, Finland for his help and support in conducting this study.

## References

1. Ries, E.: The Lean Startup: How Today's Entrepreneurs Use Continuous Innovation to Create Radically Successful Businesses. Crown Business (2011)
2. Giardino, C., Wang, X., Abrahamsson, P.: Why early-stage software startups fail: A behavioral framework. In: Software Business. Towards Continuous Value Delivery. Springer (2014) 27-41
3. Terho, H., Suonsyrjä, S., Karisalo, A., Mikkonen, T.: Ways to Cross the Rubicon: Pivoting in Software Startups. Product-Focused Software Process Improvement Volume 9459 of the series Lecture Notes in Computer Science. (2015) 555-568
4. Van der Van, J.S., Bosch, J.: Pivots and Architectural Decisions: Two Sides of the Same Medal? What Architecture Research and Lean Startup can learn from Each Other". In proceedings of Eight International Conference on Software Engineering Advances (ICSEA'13), Venice, Italy, (2013) 310-317.
5. Giardino, C., Bajwa, S.,Wang, X., Abrahamsson, P.: Key challenges in early-stage software startups. In Lassenius, C., Dingsyr, T., Paasivaara, M., eds.: Agile Processes, in Software Engineering, and Extreme Programming. Volume 212 of Lecture Notes in Business Information Processing. Springer International Publishing (2015) 52-63
6. Blank, S.: The four steps to the epiphany. Ist edn. CafePress. (2005)
7. Nazar, J.: "14 Fanous business pivots", [online] Available: http://www.forbes.com/sites/jasonnazar/2013/10/08/14 famous-business-pivots/. (Accessed: February 03, 2016). (2013)
8. Carlson, N.: "The real history of Twitter," Business Insider, [Online]. Available:http://businessinsider.com/how-twitter-was-founded-2011-4/. (Accessed: February 03, 2016) (2011)
9. Bosch, J., Holmstrm Olsson, H., Bjrk, J., Ljungblad, J.: The early stage software startup development model: A framework for operationalizing lean principles in software startups. In Fitzgerald, B., Conboy, K., Power, K., Valerdi, R., Morgan, L., Stol, K.J., eds.: Lean Enterprise Software and Systems. Vol. 167 of Lecture Notes in Business Information Processing. Springer Berlin Heidelberg (2013) 1-15
10. Nguyen-Duc, A., Seppänen, P., Abrahamsson, P.: Hunter-gatherer cycle: a conceptual model of the evolution of software startups. In Proceedings of the 2015 International Conference on Software and System Process (ICSSP 2015). NY, USA, (2015) 199-203
11. Yin, R.: Case Study Research: Design and Methods. SAGE Publications (2003)